\documentclass[aps,prl,preprint,superscriptaddress,showpacs]{revtex4-1}
\usepackage{amsmath}
\usepackage{bm}
\usepackage{graphicx}
\usepackage{xr}
\usepackage{float}
\usepackage{xcolor}
\usepackage{lipsum}

\begin{document}

\title{Acoustic manipulation through zero-thickness perforated plane with strong-coupling effects}



\author{Ming-Hao Liu}
\affiliation{Key Laboratory of Modern Acoustics, MOE, Institute of Acoustics, Department of Physics, Collaborative Innovation Center of Advanced Microstructures, Nanjing University, Nanjing 210093, People's Republic of China}


\author{Xin-Ye Zou}
\email{xyzou@nju.edu.cn}
\affiliation{Key Laboratory of Modern Acoustics, MOE, Institute of Acoustics, Department of Physics, Collaborative Innovation Center of Advanced Microstructures, Nanjing University, Nanjing 210093, People's Republic of China}
\affiliation{State Key Laboratory of Acoustics, Chinese Academy of Sciences, Beijing 100190, People's Republic of China}

\author{Bin Liang}
\affiliation{Key Laboratory of Modern Acoustics, MOE, Institute of Acoustics, Department of Physics, Collaborative Innovation Center of Advanced Microstructures, Nanjing University, Nanjing 210093, People's Republic of China}

\author{Jian-Chun Cheng}
\email{jccheng@nju.edu.cn}
\affiliation{Key Laboratory of Modern Acoustics, MOE, Institute of Acoustics, Department of Physics, Collaborative Innovation Center of Advanced Microstructures, Nanjing University, Nanjing 210093, People's Republic of China}
\affiliation{State Key Laboratory of Acoustics, Chinese Academy of Sciences, Beijing 100190, People's Republic of China}

\date{\today}

\begin{abstract}
    How to manipulate acoustic waves through thinner structures is always a challenging problem due to the linear proportional relationship between the structural thickness and the acoustic wavelength. Here, we show the possibility of breaking this relationship by the strong-coupling effects of the radiated waves on the zero-thickness two-dimensional perforated plane, rather than reducing the thickness of the three-dimensional structure by the resonance mechanism of the cavity structure. The strong-coupling effects can be achieved and regulated by the self and mutual radiation between acoustic waves from different holes in the zero-thickness plane. We experimentally demonstrate the effectiveness of our approach by implementing acoustic focusing and holography. Our work introduces a different perspective for manipulating acoustic waves and will enable the application of ultrathin acoustic devices. 
\end{abstract}


\maketitle

\section{}
    How to manipulate acoustic waves through artificial structures has been a research topic for centuries. 
    In general, when the acoustic wavelength gets longer, a thicker structure is needed to ensure the effectiveness of acoustic wave manipulation.
    Therefore, it is still difficult to reduce the size of the structure when manipulating long-wavelength acoustic waves in engineering practices \cite{morse1986theoretical}. 
    In the past decades, great progresses have been made in manipulating acoustic waves through novel structures, such as phononic crystals \cite{sigalas1993band,garcia2000theory,yang2004focusing}, 
    acoustic metamaterials \cite{liu2000locally,kaina2015negative,liu2009broadband,ma2016acoustic,zhu2011holey,cummer2016controlling,christensen2012anisotropic} 
    and acoustic metasurface \cite{ma2014acoustic,zhao2013redirection}. 
    Especially for low-frequency operations, many works have achieved excellent results through certain subwavelength resonance structures, such as Helmholtz resonators (HRs) \cite{li2015metascreen,fang2006ultrasonic,lemoult2011acoustic,zhang2009focusing}, 
    space-coiling (SC) and labyrinthine structures \cite{liang2012extreme,li2013reflected,xie2014wavefront}. 
    The thickness of the corresponding acoustic structures can even be less than 1/100 of the manipulated wavelength \cite{li2016acoustic,tang2018acoustic}. 
    Nevertheless, since the acoustic artificial structures are based on the three-dimensional structure designs to control the losses and delays of acoustic waves passing through them, even for the above resonance structures, the structure's scale must match the wavelength to maintain the resonance effect. Therefore, due to the linear proportional relationship between the structure thickness and the wavelength, the structural thickness must be increased in proportion to the wavelength to maintain its manipulation function. In this letter, we propose a method to manipulate acoustic waves by the strong-coupling effects of the self and mutual radiation (SMR) between acoustic waves from different holes in a zero-thickness two-dimensional plane, which could break the linear proportional relationship between the structure thickness and the wavelength.

    Considering the perspective of specific physical mechanisms, the essence of acoustic wave manipulation is to obtain the required air vibration velocity field on the transmission or reflection surface of the structure, and then the target acoustic field can be established by the acoustic waves radiated from the surface. Taking the above mentioned acoustic metamaterials as an example, the amplitude and phase distribution of the desired air vibration velocity field can be obtained on the exit surface due to the losses and delays caused by the resonance structures, which is always considered as the result of the acoustic wave manipulation based on acoustic structures. At the same time, it is worth noting that the acoustic radiation itself can also modulate the amplitude and phase distribution of the air vibration velocity on the exit surface by attaching an additional acoustic impedance when considering the SMR on the transmission or reflection surface. However, compared with the manipulation effect of acoustic structures, the SMR effect is always an infinitesimal and hardly affects the result of the wave manipulation through structures. Here, we carefully consider the possibility to manipulate acoustic waves only by SMR effect. Obviously, if the thickness of the acoustic structure is reduced to zero to form a perforated plane, the manipulation effect of resonance structures that depends on the existence of the inner space of the structure will disappear, but the SMR effect may still exist. Then we have the possibility to design a zero-thickness plane (ZTP) to control acoustic waves only by the SMR effect, instead of using the inner space of a traditional acoustic artificial structure, which means that the thickness of the plane will always remain zero when the wavelength get longer.

\begin{figure}[htb]
        \includegraphics[width=0.9\linewidth]{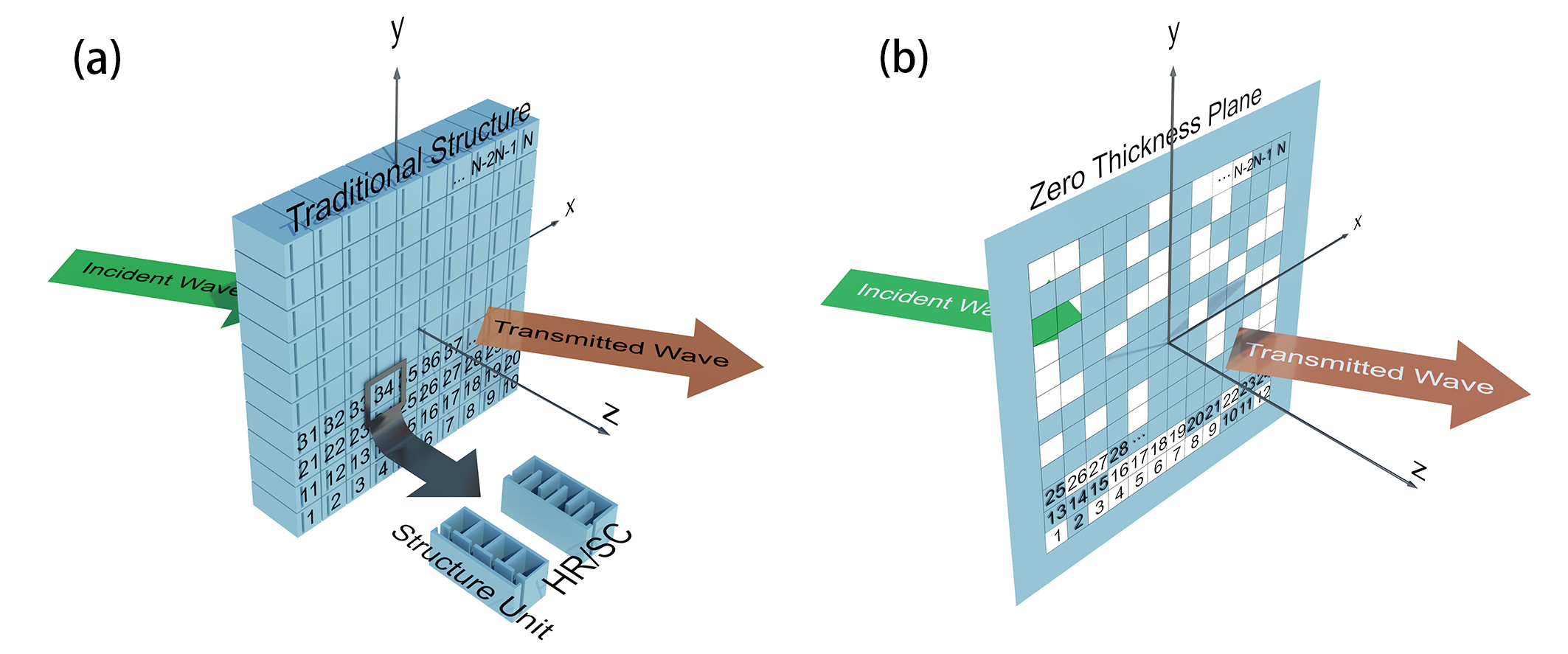}
        \caption{
            \label{fig:illustration}
            Schematic of manipulating acoustic waves through acoustic structure and perforated ZTP. (a) Artificial acoustic structure manipulates acoustic waves through different structure units, such as HR and SC structures. (b) Perforated ZTP in which the white and blue units represent a through-hole and an acoustic hard boundary, respectively.
        }
    \end{figure}

    In the following, we present a theoretical model for manipulating transmitted waves only by   controlling the strong-coupling SMR effect through a rigid perforated ZTP.
    As we known, acoustic metamaterials or phononic crystals are composed by structure units in a certain order or period, and the air vibration velocity field on the exit surface is dominated by the incident wave and the structural parameters of every unit. In order to facilitate theoretical research and comparative analysis, we number all the units by a sequence of $1, 2, 3, \cdots, I, \cdots, N$ in Fig. \ref{fig:illustration}(a), and the normal vibration velocities of the incident wave and transmitted wave of the $I^{th}$ unit are $u_i^{(I)}$ and $u_t^{(I)}$ , respectively. Then the relationship between the two normal velocities can be described as: 
    \begin{equation}
        \bm{u}_t = ( \bar{ \bm{D} }+\bar{ \bm{X}} ) \bm{u}_i,
        \label{eqn:DX}
    \end{equation}
    where $\bm{u}_i = (u_i^{(1)},u_i^{(2)},\cdots,u_i^{(I)},\cdots,u_i^{(N)})^T$ and $\bm{u}_t = (u_t^{(1)},u_t^{(2)},\cdots,u_t^{(I)},\cdots,u_t^{(N)})^T$ are two column vectors that represent normal incident and transmitted velocities of each unit. $ \bar{\bm{D}} = diag(A^{(1)}e^{j\Phi^{(1)}}, A^{(2)}e^{j\Phi^{(2)}},\cdots,A^{(I)}e^{j\Phi^{(I)}},\cdots,A^{(N)}e^{j\Phi^{(N)}}) $ is a macroscopic description of the structure's ability to manipulate acoustic waves, whose element $ A^{(I)} e^{j\Phi^{(I)}} $ represents the amplitude gain and phase delay caused by each unit. $\bar{\bm{X}}$ is a matrix that denotes a macroscopic description of the SMR effect on both reflection and transmission surfaces, whose diagonal and non-diagonal elements express the self radiation effect and the mutual radiation effect between different units, respectively. As mentioned above, the SMR effect $\bar{\bm{X}}$ is much smaller than the manipulation function $\bar{\bm{D}}$ for most acoustic structures. So the $\bar{\bm{X}}$ can be ignored without affecting the manipulation result, and Eq. (\ref{eqn:DX}) becomes the general form:
    \begin{equation}
        \bm{u}_t = \bar{ \bm{D} } \bm{u}_i.
    \end{equation}

    However, if we compress the acoustic structure into a perforated ZTP as shown in Fig. \ref{fig:illustration}(b), the manipulation function $\bar{\bm{D}}$ of the structure will disappear when the structure disappears, and the acoustic wave manipulation is possible dominated by the SMR effect $\bar{\bm{X}}$ from the perforated plane. Then Eq. (\ref{eqn:DX}) becomes:
    \begin{equation}
        \bm{u}_t = \bar{ \bm{X} } \bm{u}_i.
    \end{equation}

    So far, the acoustic manipulation problem has become how to implement a specific matrix $\bar{\bm{X}}$ to produce the desired transmitted acoustic field. Specifically, the SMR effect $\bar{\bm{X}}$ is contributed by acoustic radiation forces of the radiated acoustic fields from all units on the perforated ZTP. For each through-hole unit on the plane in Fig. \ref{fig:illustration}(b), the air inside the hole could be considered as an ultrathin air plate that will be forced to vibrate by the incident, reflected and transmitted acoustic waves. Since the scale of each hole is much smaller than the wavelength, the radiated wave from each air plate can be regarded as a spherical wave propagating along the surface. The radiated spherical wave will give a force to the air plate in each hole and change their vibration state. Thus, all these radiation forces will interact on all the units and modulate the air velocity distribution on the plane. Because there is no acoustic structure but only a perforated plane, the SMR effect $\bar{\bm{X}}$ will become the strong-coupling effect and the main means of manipulating acoustic waves.

    Although the ZTP can be theoretically designed as any curved profile, we use a square rigid plane as an example in our research in order to facilitate theoretical calculation and experimental measurement. The square rigid plane is an acoustic hard boundary and is fixed in an unbounded space such as an anechoic chamber. We carve pattern by drilling holes on the plane to allow acoustic waves to pass through, so that we can adjust the SMR effect and achieve the required air vibration velocity distribution on the plane. The pattern area is sequentially divided into N small square units with the same size in Fig. \ref{fig:illustration}(b), and each unit could be a through-hole (white) or a hard boundary (blue). The pattern area is described by a vector $ \bm{v} = (1,1,0,\cdots,0) $ whose $I^{th}$ element represents the state of $I^{th}$ unit on the pattern: a through-hole (1) or a hard boundary (0). Finally, the distribution of the vibration velocity on the transmission surface can be obtained as: 
    \begin{equation}
        \bm{u}_t =
        \frac{1}{2}
        (
            \bm{E}
            +
            \bm{z}_{xr}^{-1} \bm{R}_0
        )
        \bm{u}_i,
    \end{equation}
    where $\bm{E}$ is an N-dimensional unit matrix, $\bm{R}_0$ is an N-dimensional diagonal matrix whose $I^{th}$ diagonal element denotes the characteristic impedance of the incident wave at $I^{th}$ unit on the plane. $\bm{z}_{xr}$ is a matrix that represents the SMR effect of the pattern drilled on the plane, the matrix element at $(I,J)$ position expresses the self radiation effect $(I=J)$ and the mutual radiation effect $(I\neq J)$: $z_{xr}^{(I)\leftarrow(J)} = jk\rho_0 c_0 g(\bm{r}_I;\bm{r}_J) S_0$, where $j$ is the imagnary unit, $\rho_0$ and $c_0$ are the density and the acoustic speed of the media, $S_0$ is the area of per unit on the plane, $\bm{r}_I$, $\bm{r}_J$ are the location vectors of $I^{th}$ and $J^{th}$ units and $g(\bm{r}_I;\bm{r}_J)$ is the Green's function between these two positions. If the $I^{th}$ unit is an acoustic hard boundary but not a hole, the  $I^{th}$ row in $\bm{z}_{xr}$ should be rewritten into a vector as $(0,0,0,\cdots,-1,\cdots,0)$, in which -1 is only located on the diagonal of the matrix and the others are zero. More theory details including the reflection situation are given in the supplementary materials \cite{[{See }][{ at urlXXX.}]Supplement}.
    
    Therefore, if the size of the rigid ZTP, the boundary conditions of the surrounding environment and the incident wave are preset, we can manipulate acoustic waves radiated from the transmission surface of the plane by designing a specific pattern drilled on it. By now, the acoustic wave manipulation process transfers from the interior of acoustic structure to the surface outside of the structure. That is, the manipulation of acoustic waves would no longer depend on the interior space of structure when we compress structure into zero, and finally the strong-coupling SMR effect $\bar{\bm{X}}$ can be achieved by the perforated ZTP. For a given acoustic wave manipulation task, the pattern design of the perforated ZTP becomes a mathematical optimization problem, which aims to find out a specific pattern vector $\bm{v}$ to make a minimum difference between the transmitted acoustic field and the desired field. We use the optimization algorithm method to find out the specific pattern on the ZTP for the specific acoustic application. Meanwhile, an NVIDIA graphic card is used to accelerate matrix calculation speed. The flow diagram of optimization algorithm is given in the Fig. S2 of the supplementary material \cite{[{See }][{ at urlXXX.}]Supplement}.

    In order to verify the validity and practicability of our approach, we propose two specific examples realized on the transmission side: acoustic focusing and acoustic holography. In these two cases, a square steel plate with 0.6m side length and 1mm thickness is used as the rigid ZTP. The experiment details are given in the supplementary materials \cite{[{See }][{ at urlXXX.}]Supplement}.

    \begin{figure*}[htb]
        \includegraphics[width=0.8\linewidth]{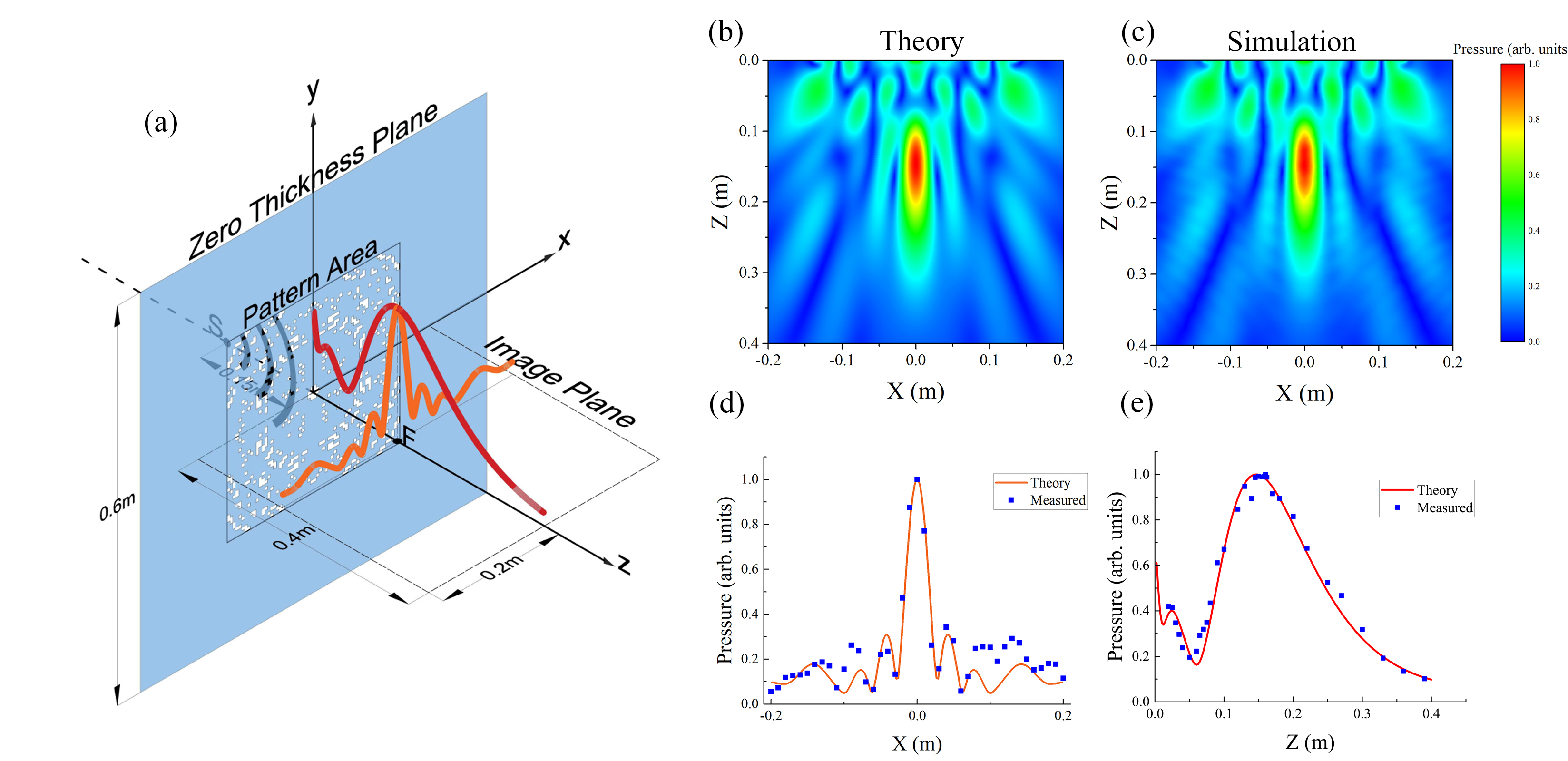}
        \caption{
            \label{fig:Resuls_focusing}
            Results of acoustic focusing. (a) Illustration of acoustic focusing. Point S is the source point and point F is the focus. (b) Theory result of acoustic field on image plane. (c) Simulation result of acoustic field on image plane. (d) Theory and experiment results on the cross-section line (orange line in (a)). (e) Theory and experiment results on the z-axis (red line in (a)). 
        }
    \end{figure*}

\begin{figure}[htb]
        \includegraphics[width=0.5\linewidth]{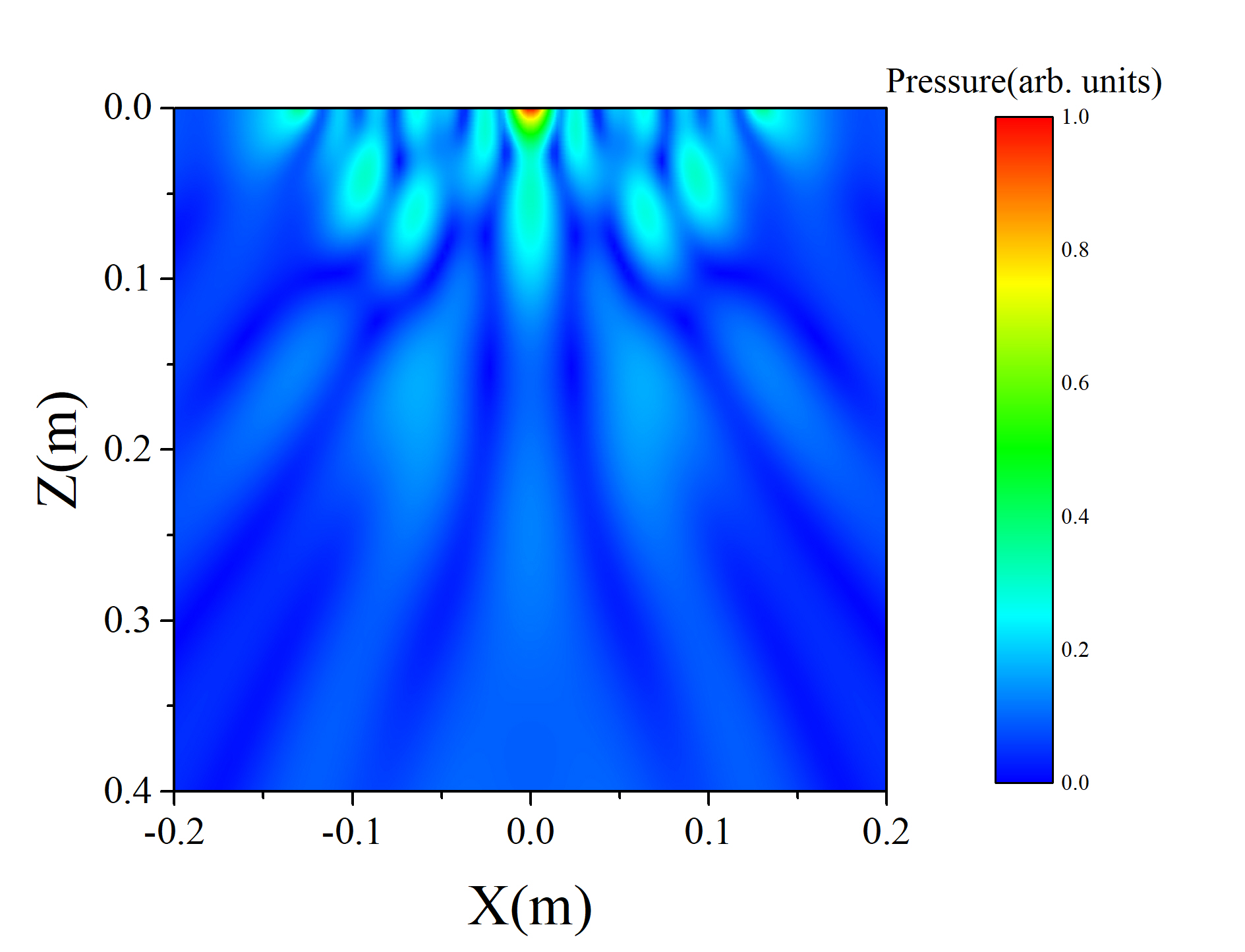}
        \caption{
            \label{fig:Result_noMR}
            The theory result of acoustic focusing without considering the SMR effect.
        }
    \end{figure}

    In the acoustic focusing example, the pattern area is the square with a side length of 0.3m. We divide each side of the pattern area into 60 parts, and then 3600 units can be obtained. The 8.5kHz incident wave is radiated by a point source, which is located at 15cm away from the steel plate on its center axis in Fig. \ref{fig:Resuls_focusing}(a). In the expected acoustic field as we design, the focus is located on the center axis and is 14.75cm away from the plate. Considering the symmetry of this problem, we designed the pattern area on four-axis symmetry (horizontal, vertical, two diagonal lines). The calculation area of acoustic field optimization is a square area perpendicular to the rigid plane, whose side length is 0.4m. Fig. \ref{fig:Resuls_focusing}(b) and Fig. \ref{fig:Resuls_focusing}(c) give the theory and simulation results of the acoustic field, respectively. We performed the corresponding experiment in the anechoic chamber, and the acoustic fields measured on the cross-section line and the focal axis are shown in Fig. \ref{fig:Resuls_focusing}(d) and Fig. \ref{fig:Resuls_focusing}(e). The experimental result agrees well with the theoretical result, which shows that the rigid perforated plane works well in acoustic focusing situation. Meanwhile, the theory result of acoustic focusing without considering the SMR effect is also given in Fig. \ref{fig:Result_noMR}, which shows that the acoustic focusing phenomenon disappears without considering the SMR effect. In this theory calculation, the non-diagonal elements in $\bm{z}_{xr}$ is set into 0, that means, we do not consider the mutual radiation effect between units on the plane. All other parameters in the calculation are set as same as the acoustic focus example in Fig. \ref{fig:Resuls_focusing}. Comparing Fig. \ref{fig:Resuls_focusing}(b) and Fig. \ref{fig:Result_noMR}, we can find that the SMR effect is the strong-coupling effects and the most important factor to manipulate acoustic waves for the rigid perforated ZTP.

\begin{figure*}[htb]
        \includegraphics[width=0.8\linewidth]{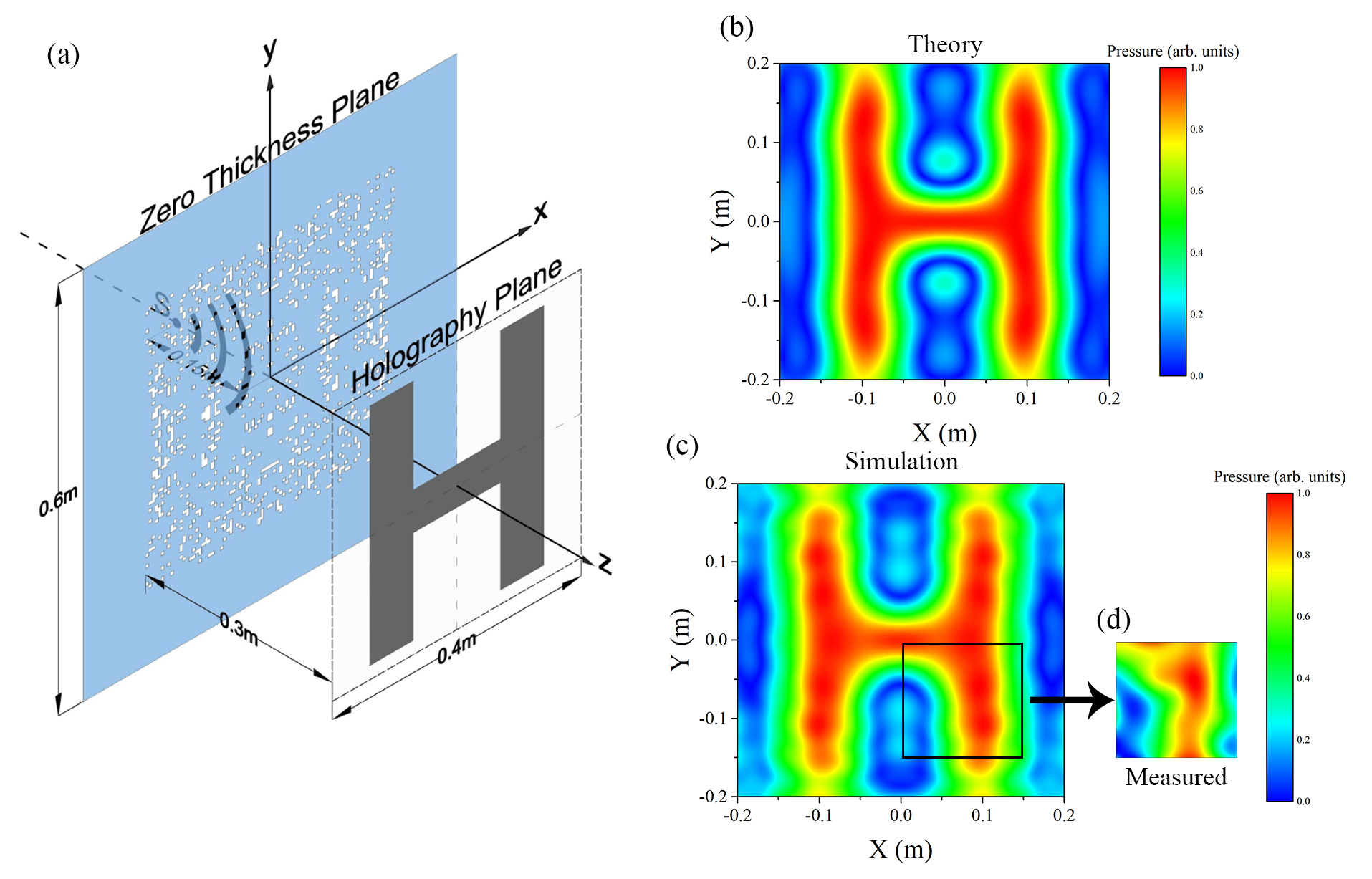}
        \caption{
            \label{fig:Result_Holo}
            Results of acoustic holography. (a) Illustration of the experiment. A letter `H' is designed as the desired acoustic field on the holography image plane. (b) Theoretical acoustic field on the image plane. (c) Simulation result on the image plane. (d) Measurement result in experiment. About 1/4 area of the holography plane is measured because of the symmetry.
        }
    \end{figure*}

    Furthermore, in order to verify the method's flexibility, we design another rigid perforated plane to realize acoustic holography which is to radiate specific acoustic field into space \cite{melde2016holograms,tian2017acoustic,zhu2018fine}. As same as the acoustic focusing example, a square steel plate is also used as the rigid ZTP. The pattern area with a side length of 0.4m has been divided into 6400 units with 80 pieces on each side. The incident acoustic wave is radiated by a point source, which is 15cm far away from the plane and is located on the center axis of the plane. The image plane is parallelized to the steel plate and the distance between them is 0.3m. The desired acoustic field on the transmission side is a letter `H' on the image plane, whose side length is 0.4m in Fig. \ref{fig:Result_Holo}. The theory and simulation results of the transmitted acoustic field on the image plane are shown in Fig. \ref{fig:Result_Holo}(b) and Fig. \ref{fig:Result_Holo}(c). Since there is two axis symmetry (horizontal and vertical) in this problem, we just need to measure 1/4 part of the image plane in Fig. \ref{fig:Result_Holo}(d). It can be found that the measurement result fits well with the corresponding simulation and theory results. So far, from the above acoustic focusing and holography examples, we can achieve the conclusion that the SMR effect will be the strong-coupling effects for the rigid perforated ZTP, which can be successfully applied for different acoustic applications. Our method is also suitable for manipulating reflected waves, the acoustic holography examples of letters imaging and panda imaging on the reflection side of the ZTP are given in the supplementary materials Fig. S4 and Fig. S5 \cite{[{See }][{ at urlXXX.}]Supplement}.

    In conclusion, we have demonstrated that acoustic waves radiate outward from the exit surface of a structure can be regarded not only as the result of wave manipulation but also as an efficient way to manipulate acoustic waves. We propose a new perspective and method to manipulate acoustic wave propagation by the strong-coupling SMR effect through a rigid perforated ZTP. It transforms the design problem of the three-dimensional functional acoustic material into the perforation design in the two-dimensional rigid plane. Ultrathin acoustic devices can be designed in practical engineering applications, especially for low-frequency acoustic waves.

\begin{acknowledgments}
    This work was supported by the National Key R\&D Program of China, (grant no. 2017YFA0303700), National Natural Science Foundation of China (grant nos. 11634006, 81127901, and 11690030), State Key Laboratory of Acoustics, Chinese Academy of Sciences.

	Ming-Hao Liu and Xin-Ye Zou contributed equally to this article.

\end{acknowledgments}

\bibliography{Ref}

\end{document}


\title{Supplementary Materials}

\date{\today}
\maketitle

\section{Theoretical derivation}
\begin{figure}[htb]
        \includegraphics[]{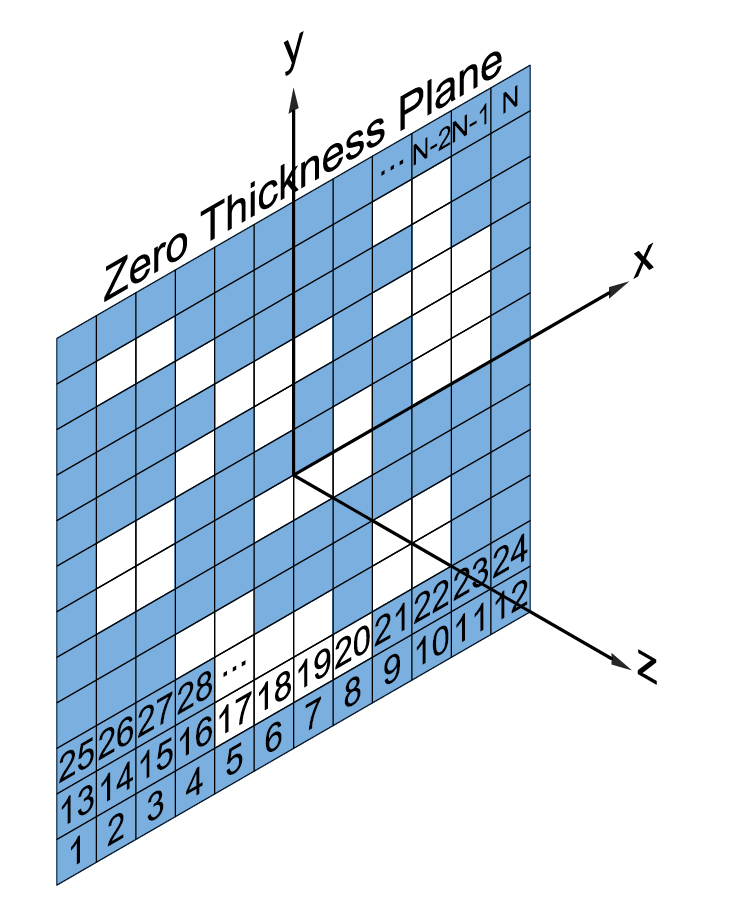}
        \caption{
            \label{fig:ZIPstr}
            The illustration of ZTP. The rigid plane is divided into N units, and we marked all the units in a sequence $1, 2, \cdots, N$. The blue unit represents the acoustic hard boundary (like a wall) that totally reflects the incident wave, and the white unit represents the through-hole that allow the incident wave to pass through.
        }
    \end{figure}

    Here, we deduce the theoretical result of the perforated zero-thickness plane (ZTP). We assume the ZTP as a rigid square plate which is located in an unbounded free space. The origin of the Cartesian coordinate system is at the center of the ZTP, where the z-axis is perpendicular to the plate. The plate is divided into N small square units of the same size, the $I^{th}$ unit is recorded as $\Sigma^{(I)}$, and the scale of every unit is much smaller than the wavelength (Fig. \ref{fig:ZIPstr}).

    For every unit on the ZTP, we denote the set of all the acoustic hard boundary units as $\Sigma_w$, and the  set of all the through-hole units as $\Sigma_t$, respectively. If the $I^{th}$ unit is an acoustic hard boundary, that means $\Sigma^{(I)}\in\Sigma_w$; similarly, if the $I^{th}$ unit is a hole, then $\Sigma^{(I)}\in\Sigma_t$.

    If the $I^{th}$ unit is an acoustic hard boundary, the normal vibration velocity of the acoustic wave must be zero on both sides of this unit:
    \begin{equation}
        \left\{
            \begin{aligned}
                \vec{u}_i^{(I)} \cdot \vec{n} + 
                \sum_{J=1}^N{\vec{u}_r^{(I)\leftarrow(J)}\cdot \vec{n}} 
                &= 0, &\Sigma^{(I)}\in\Sigma_w, \\
                \vec{u}_t^{(I)} \cdot \vec{n} &= 0, &\Sigma^{(I)}\in\Sigma_w,
            \end{aligned}
        \right.
        \label{eqn:CondW}
    \end{equation}
    where $\vec{n}$ is the unit vector of the normal direction of the plane, $\vec{u}_i^{(I)}$ and $\vec{u}_t^{(I)}$ are the vibration velocities of incident and transmitted wave at the $I^{th}$ unit, $\vec{u}_r^{(I)\leftarrow(J)}$ means the vibration velocity of the reflected wave that radiated from the $J^{th}$ unit to the $I^{th}$ unit.

    If the $I^{th}$ unit is a through-hole, the acoustic boundary condition on it is the continuity of the normal vibration velocity and the acoustic pressure:
    \begin{equation}
        \left\{
            \begin{aligned}
                \vec{u}_i^{(I)} \cdot \vec{n} + 
                \sum_{J=1}^N{\vec{u}_r^{(I)\leftarrow(J)}\cdot \vec{n}} 
                &= 
                \sum_{J=1}^N{\vec{u}_t^{(I)\leftarrow(J)}\cdot \vec{n}}, &\Sigma^{(I)}\in\Sigma_t, \\
                p_i^{(I)} + 
                \sum_{J=1}^{N}{p_r^{(I)\leftarrow(J)}} 
                &=
                \sum_{J=1}^{N}{p_t^{(I)\leftarrow(J)}}, &\Sigma^{(I)}\in\Sigma_t,
            \end{aligned}
        \right.
        \label{eqn:CondT}
    \end{equation}
    where $\vec{u}_t^{(I)\leftarrow(J)}$ means the vibration velocity of the transmitted wave that radiated from the $J^{th}$ unit to the $I^{th}$ unit, $p_i^{(I)}$ is the incident acoustic pressure on the $I^{th}$ unit, $p_r^{(I)\leftarrow(J)}$ and $p_t^{(I)\leftarrow(J)}$ are the reflected and transmitted acoustic pressure that radiated from the $J^{th}$ unit to the $I^{th}$ unit, respectively.

    It is worth noting that the incident wave can be a plane wave, a spherical or any other type acoustic wave. Meanwhile, we suppose that the wavelength of the incident wave is smaller than the side length of the ZTP, so the diffraction effect could be ignored here. We also assumed that the scale of each unit is much smaller than the wavelength, then the radiated acoustic field caused by each unit (including reflected and transmitted field) can be regarded as a spherical wave radiated from a point source. So on the reflection or transmission surface, the vibration velocity vector of the acoustic wave radiating from one unit to another is parallel to the rigid plane:

    \begin{equation}
        \vec{u}_r^{(I)\leftarrow(J)} \cdot \vec{n} = 0, \quad
        \vec{u}_t^{(I)\leftarrow(J)} \cdot \vec{n} = 0  \quad 
        (I \neq J).
        \label{eqn:u_parallel_is_zero}
    \end{equation}

    Considering Eq. (\ref{eqn:u_parallel_is_zero}), we can replace $\vec{u}_i^{(I)} \cdot \vec{n}$, $\vec{u}_r^{(I)\leftarrow(I)}\cdot \vec{n}$, $\vec{u}_t^{(I)\leftarrow(I)}\cdot \vec{n}$ with ${u}_i^{(I)}$, ${u}_r^{(I)}$, ${u}_t^{(I)}$, then Eq. (\ref{eqn:CondW}) and (\ref{eqn:CondT}) can be written as:

    \begin{equation}
        \left\{
            \begin{aligned}
                {u}_i^{(I)} + {u}_r^{(I)} &= 0, &\Sigma^{(I)}\in\Sigma_w, \\
                {u}_t^{(I)} &= 0, &\Sigma^{(I)}\in\Sigma_w,
            \end{aligned}
        \right.
        \label{eqn:CondW_simplified}
    \end{equation}

    \begin{equation}
        \left\{
            \begin{aligned}
                {u}_i^{(I)} + {u}_r^{(I)} &= {u}_t^{(I)}, &\Sigma^{(I)}\in\Sigma_t, \\
                p_i^{(I)} + 
                \sum_{J=1}^{N}{p_r^{(I)\leftarrow(J)}} 
                &=
                \sum_{J=1}^{N}{p_t^{(I)\leftarrow(J)}}, &\Sigma^{(I)}\in\Sigma_t.
            \end{aligned}
        \right.
        \label{eqn:CondT1}
    \end{equation}

    Next, the relationship between the acoustic pressure of the radiated wave and the vibration velocity from its source unit can be given as:

    \begin{equation}
        \left\{
            \begin{aligned}
                p_r^{(I)\leftarrow(J)} &= -jk\rho_0c_0 u_r^{(I)} g(\vec{r}^{(I)}; \vec{r}^{(J)}) S_0, \\
                p_t^{(I)\leftarrow(J)} &=  jk\rho_0c_0 u_t^{(I)} g(\vec{r}^{(I)}; \vec{r}^{(J)}) S_0,
            \end{aligned}
        \right.
        \label{eqn:source_radiation}
    \end{equation}
    where $S_0$, $k$, $\rho_0$, $c_0$, $\vec{r}^{(I)}$, $\vec{r}^{(J)}$, $g(\vec{r}^{(I)}; \vec{r}^{(J)})$ are the area of the unit, the wave number, the density of the background media, the sound speed of the background media, the position vector of the $I^{th}$ and the $J^{th}$ unit, and the Green's function between these two locations: $\vec{r}^{(I)}$ and $\vec{r}^{(J)}$ , respectively. We use semi-unbounded free space Green's function as an approximation:

    \begin{equation}
        g(\vec{r}^{(I)};\vec{r}^{(J)}) = 
        \frac{
            e^{ -jk(\vec{r}^{(I)}-\vec{r}^{(J)}) }
        }{
            2\pi | \vec{r}^{(I)}-\vec{r}^{(J)} |
        }.
    \end{equation}

    Here, we define 
    $z_{xr}^{(I)\leftarrow(J)} = -jk\rho_0c_0 g(\vec{r}^{(I)}; \vec{r}^{(J)}) S_0 $ 
    to express the impedance relationship in Eq. (\ref{eqn:source_radiation}) as:

    \begin{equation}
        \left\{
            \begin{aligned}
                p_r^{(I)\leftarrow(J)} &= - z_{xr}^{(I)\leftarrow(J)} u_r^{(I)}, \\
                p_t^{(I)\leftarrow(J)} &=   z_{xr}^{(I)\leftarrow(J)} u_t^{(I)}.
            \end{aligned}
        \right.
        \label{eqn:impdance_relationship}
    \end{equation}

    Then substituting Eq. (\ref{eqn:impdance_relationship}) into Eq. (\ref{eqn:CondT1}), we can obtain:

    \begin{equation}
        \left\{
            \begin{aligned}
                {u}_i^{(I)} + {u}_r^{(I)} &= {u}_t^{(I)}, &\Sigma^{(I)}\in\Sigma_t, \\
                R_0^{(I)} u_i^{(I)} + 
                \sum_{J=1}^{N}{-z_{xr}^{(I)\leftarrow(J)} u_r^{(J)} } 
                &=
                \sum_{J=1}^{N}{ z_{xr}^{(I)\leftarrow(J)} u_t^{(J)} }, &\Sigma^{(I)}\in\Sigma_t,
            \end{aligned}
        \right.
        \label{eqn:CondT2}
    \end{equation}
    where  $R_0^{(I)}$ is the characteristic impedance of incident wave on the $I^{th}$ unit, which is the ratio of acoustic pressure to normal vibration velocity. Then, from Eq. (\ref{eqn:CondW_simplified}) and (\ref{eqn:CondT2}), we can obtain the equations that describe the whole ZTP:

    \begin{equation}
        \bm{u}_i + \bm{u}_r = \bm{u}_t,
        \label{eqn:final_u}
    \end{equation}

    \begin{equation}
        \begin{aligned}
            \left(\begin{matrix}
                R_0^{(1)} & & & & & \\
                & R_0^{(2)} & & & & \\
                & & \ddots & & & \\
                & & & \phantom{2} 1\phantom{2} & & \\
                & & & & \ddots & \\
                & & & & & R_0^{(N)}
            \end{matrix}\right)
            \left(\begin{matrix}
                u_i^{(1)} \\
                u_i^{(2)} \\
                \vdots \\
                u_i^{(I)} \\
                \vdots \\
                u_i^{(N)}
            \end{matrix}\right)
            -
            &\left(\begin{matrix}
                z_{xr}^{(1)\leftarrow(1)} &z_{xr}^{(1)\leftarrow(2)} &\cdots &z_{xr}^{(1)\leftarrow(N)} \\
                z_{xr}^{(2)\leftarrow(1)} &z_{xr}^{(2)\leftarrow(2)} &\cdots &z_{xr}^{(2)\leftarrow(N)} \\
                \vdots &\vdots &\ddots &\vdots \\
                0 &0 &-1 &0 \\
                \vdots &\vdots &\ddots &\vdots \\
                z_{xr}^{(N)\leftarrow(1)} &z_{xr}^{(N)\leftarrow(2)} &\cdots &z_{xr}^{(N)\leftarrow(N)} \\
            \end{matrix}\right)
            \left(\begin{matrix}
                u_r^{(1)} \\
                u_r^{(2)} \\
                \vdots \\
                u_r^{(I)} \\
                \vdots \\
                u_r^{(N)}
            \end{matrix}\right) \\
            =
            &\left(\begin{matrix}
                z_{xr}^{(1)\leftarrow(1)} &z_{xr}^{(1)\leftarrow(2)} &\cdots &z_{xr}^{(1)\leftarrow(N)} \\
                z_{xr}^{(2)\leftarrow(1)} &z_{xr}^{(2)\leftarrow(2)} &\cdots &z_{xr}^{(2)\leftarrow(N)} \\
                \vdots &\vdots &\ddots &\vdots \\
                0 &0 &-1 &0 \\
                \vdots &\vdots &\ddots &\vdots \\
                z_{xr}^{(N)\leftarrow(1)} &z_{xr}^{(N)\leftarrow(2)} &\cdots &z_{xr}^{(N)\leftarrow(N)} \\
            \end{matrix}\right)
            \left(\begin{matrix}
                u_t^{(1)} \\
                u_t^{(2)} \\
                \vdots \\
                0 \\
                \vdots \\
                u_t^{(N)}
            \end{matrix}\right).
        \end{aligned}
        \label{eqn:final_p_comp}
    \end{equation}

    The simplified form of Eq. (\ref{eqn:final_p_comp}) is:
    \begin{equation}
        \bm{R}_0 \bm{u}_i- \bm{z}_{xr} \bm{u}_r = \bm{z}_{xr} \bm{u}_t,
        \label{eqn:final_p}
    \end{equation}
    where $ \bm{u}_i = (u_i^{(1)}, u_i^{(2)}, \cdots, u_i^{(N)})^T $, $ \bm{u}_r = (u_r^{(1)}, u_r^{(2)}, \cdots, u_r^{(N)})^T $, $ \bm{u}_t = (u_t^{(1)}, u_r^{(2)}, \cdots, u_t^{(N)})^T $ are three vectors that describe the set of normal vibration velocity on all the units of the incident, reflected and transmitted waves, respectively. $ \bm{R}_0 = \diag( {R}_0^{(1)}, {R}_0^{(2)}, \cdots, {R}_0^{(N)}) $ is a diagonal matrix that expresses the characteristic impedance of the incident wave on all the units, and $ \bm{z}_{xr} $ is a matrix that represents the self and mutual radiation (SMR) effect between all the units. The matrix unit of $ \bm{z}_{xr} $ is:

    \begin{equation}
        \bm{z}_{xr} (I,J) = \left\{
        \begin{aligned}
            z_{xr}^{(I)\leftarrow(J)}, \quad &\Sigma^{(I)}\in \Sigma_t, \\
             0, \quad &\Sigma^{(I)}\in \Sigma_w \ &(I\neq J), \\
            -1, \quad &\Sigma^{(I)}\in \Sigma_t \ &(I = J).
        \end{aligned}
        \right.
    \end{equation}

    Eq. (\ref{eqn:final_u}) and (\ref{eqn:final_p}) give the math descriptions of the ZTP. Since $ \bm{z}_{xr} $ is reversible for most situations, from these two equations, we get:
    \begin{equation}
        \left\{
            \begin{aligned}
                \bm{u}_t &= \frac{1}{2} 
                (
                    \bm{E} + \bm{z}_{xr}^{-1} \bm{R}_0
                ) \bm{u}_i, \\
                \bm{u}_r &= \frac{1}{2} 
                (
                    -\bm{E} + \bm{z}_{xr}^{-1} \bm{R}_0
                ) \bm{u}_i,
            \end{aligned}
        \right.
        \label{eqn:final_all}
    \end{equation}
    where $\bm{E}$ is the unit matrix. So, if we have the specific pattern on the ZTP ($\bm{z}_{xr}$) and the details of incident wave ($\bm{u}_i$ and $\bm{R}_0$), we can get $\bm{u}_t$ or $\bm{u}_r$ from Eq. (\ref{eqn:final_all}), and we can obtain the reflected acoustic field and transmitted acoustic field in the end.

\pagebreak

\section{Flow diagram of optimization algorithm}
    \begin{figure}[htb]
        \includegraphics[width=0.8\linewidth]{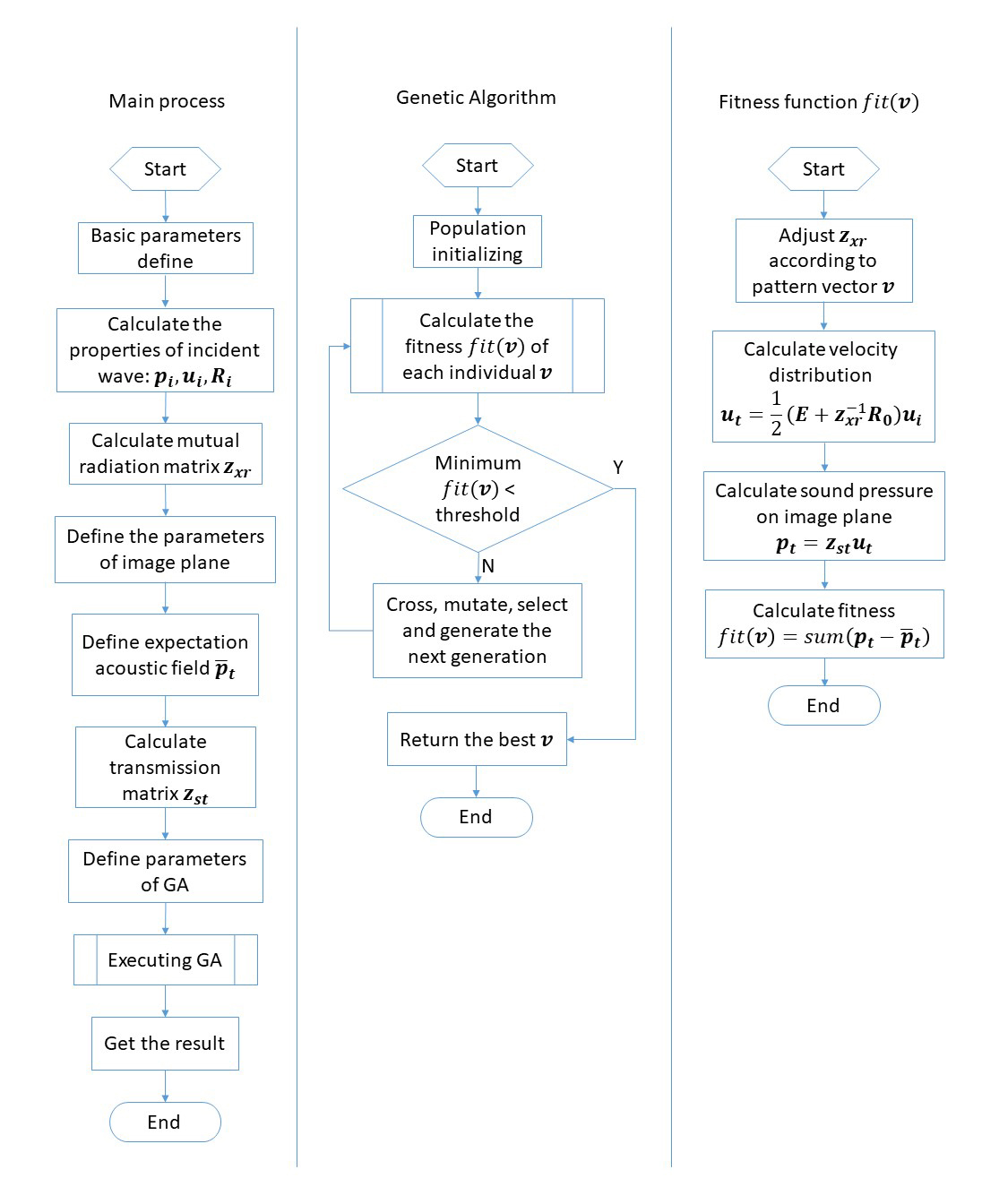}
        \caption{
            \label{fig:FlowDiag}
            The flow diagram of optimization algorithm. Genetic algorithm (GA) is used to search the optimized result of the pattern.
        }
    \end{figure}

\pagebreak

\section{Experiment details}
\begin{figure}[htb]
        \includegraphics[width=0.7\linewidth]{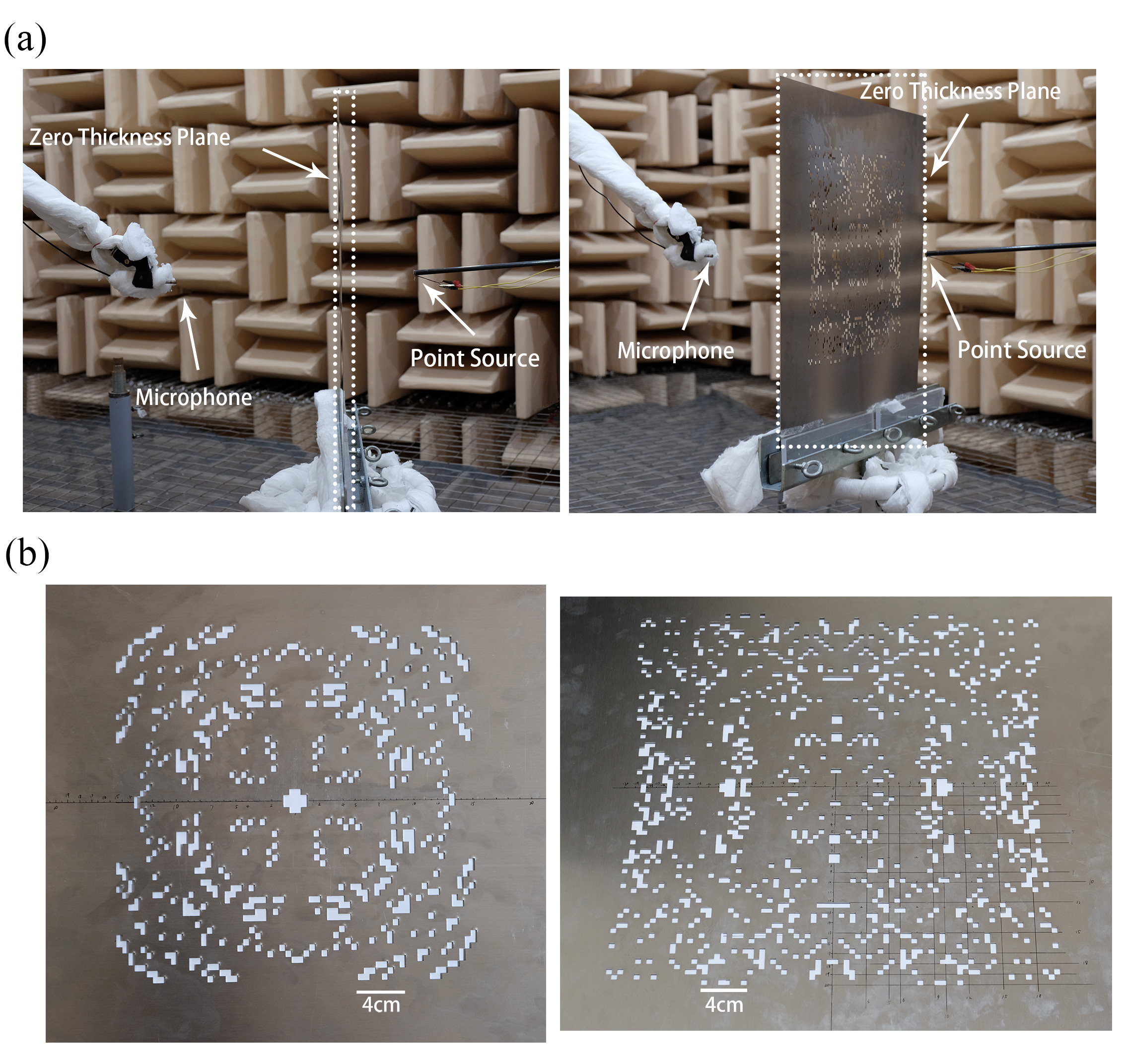}
        \caption{
            \label{fig:experiment_detail}
            Illustration of the experiment. (a) The experiment environment in the anechoic chamber. White absorbing cottons are used to control acoustic scattering. (b) The pattern drilled on the steel plane. The left is the focusing pattern and right is the holography pattern.
        }
    \end{figure}

    As shown in Fig. \ref{fig:experiment_detail}, in the corresponding experiment, the pattern for holography or focusing is drilled on a square steel plate with 0.6m side length and 1mm thickness. Considering the wavelength of the manipulated wave, the steel plate could be regarded as a rigid plane close to zero-thickness. We carry out the related experimental measurements in the anechoic chamber. The perforated ZTP is set in the center of the anechoic chamber. The acoustic point source (a piezoelectric ceramic sheet) is located on the axis of the ZTP and the microphone is fixed on the other side of the plane. The B\&K PULSE system is used as our measurement instrument, which records the acoustic pressure measured by the microphone in space.

\section{Reflected wave manipulation example}
    To further verify the manipulation ability of the ZTP, we also provide an acoustic holography example on the reflection side of the ZTP. According to Eq. (\ref{eqn:final_all}), we can obtain the desired acoustic field on the image plane by the distribution of reflection velocity $\bm{u}_r$ on the ZTP:

    \begin{equation}
        \bm{u}_r = \frac{1}{2} 
        (
            -\bm{E} + \bm{z}_{xr}^{-1} \bm{R}_0
        ) \bm{u}_i.
        \label{eqn:final_ur}
    \end{equation}

    We give three specific examples here: two are radiating letters `m' and `N' on the image plane and the third one is radiating a picture of panda on the image plane. The incident wave is a plane wave whose wave vector is vertical to the ZTP. The frequency of incident wave is 6kHz (letter `m' and `N') or 4kHz (panda). The image plane is located at the reflection side of the ZTP and 1cm away from the plane. For acoustic wave at 4kHz or 6kHz, 1cm is a near-field distance that means we can use the evanescent wave in near-field to achieve higher resolution imaging.

\begin{figure}[htb]
        \includegraphics[width=0.6\linewidth]{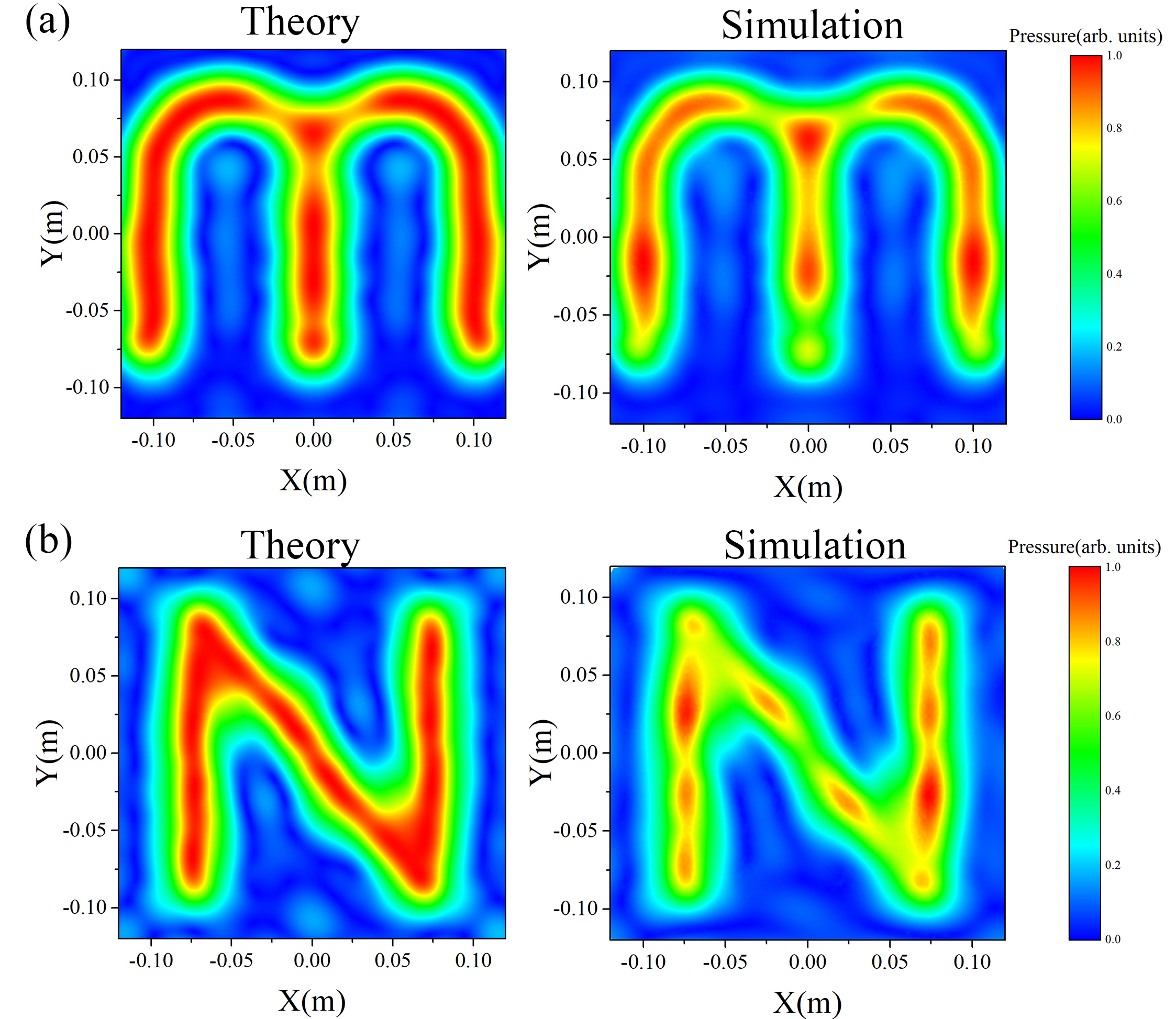}
        \caption{
            \label{fig:MN}
            Verifying the reflected wave manipulation ability of the ZTP. (a) Imaging of letter `m'. (b) Imaging of letter `N'. The result expresses that the ZTP is also effective in controlling acoustic waves on the reflection side.
        }
    \end{figure}

    In letter imaging, the ZTP is a square with a side length of 0.6m. The image plane is a square parallel to the ZTP whose side length is 0.24m. The imaging result is shown in Fig. \ref{fig:MN}.

\begin{figure}[htb]
        \includegraphics[width=\linewidth]{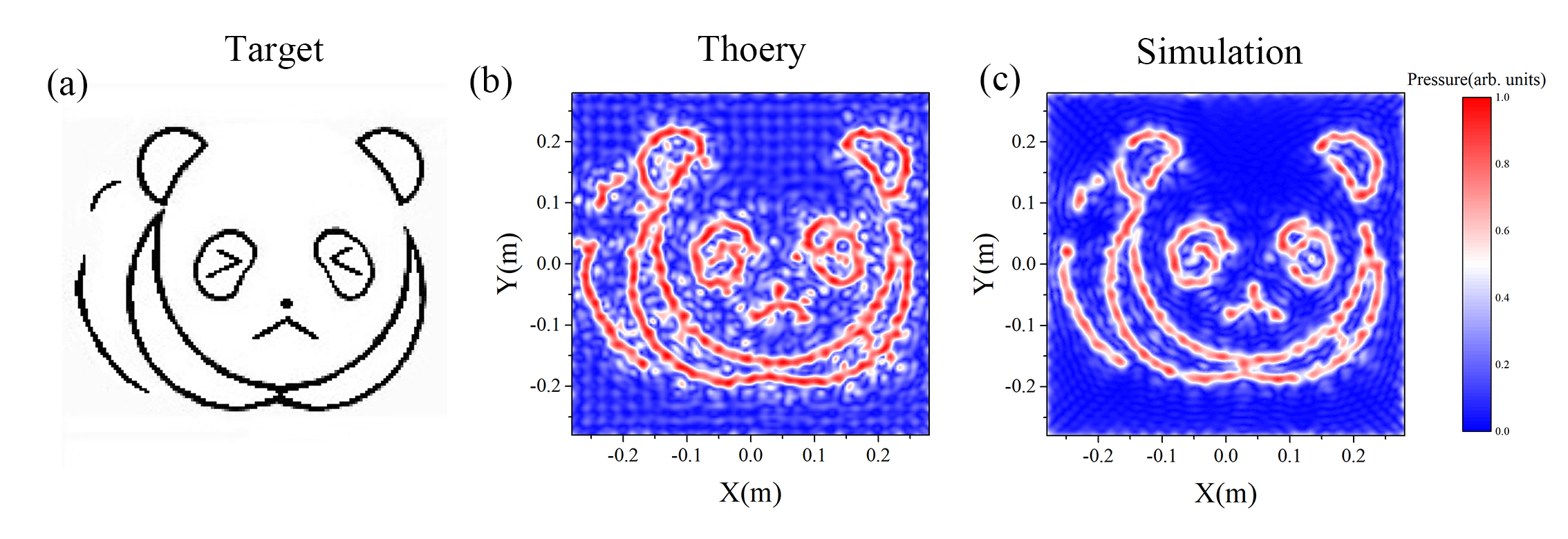}
        \caption{
            \label{fig:gungun}
            Using the ZTP to manipulates meticulous pattern. (a) The desired acoustic field: a cartoon panda. (b) Theory result. (c) Simulation result. The results show that the ZTP could also manipulate acoustic waves with evanescent wave to construct a meticulous acoustic field as desired.
        }
    \end{figure}

    In panda imaging, the ZTP and image plane are both square. The border lengths of the ZTP and image plane are 0.8m and 0.46m, respectively. The final result is expressed in Fig. \ref{fig:gungun}.
    
